\documentclass[runningheads,envcountsect,envcountsame]{llncs}

\usepackage[T1]{fontenc}
\usepackage{fixltx2e}
\usepackage{graphics}
\usepackage{srcltx}
\usepackage{amssymb,amsmath,amsfonts}
\usepackage{textcomp}
\usepackage{epsf}
\usepackage[dvips]{epsfig}
\begin{document}

\author{Julien Cassaigne\inst{1} \and Anna E. Frid\inst{1,3} \and Svetlana Puzynina\inst{2,3} \and Luca Q. Zamboni \inst{2,4}}
\institute{Aix-Marseille Universit\'{e}, France
\email{cassaigne@iml.univ-mrs.fr, anna.e.frid@gmail.com}
 \\
\and Department of Mathematics and Statistics, University of Turku, Finland
\email{svepuz@utu.fi}\\
\and Sobolev Institute of Mathematics, Russia \\
\and Universit\'e de Lyon 1, France
\email{zamboni@math.univ-lyon1.fr}}
\authorrunning{J. Cassaigne, A. E. Frid, S. Puzynina, L. Q. Zamboni}
\title{Subword complexity and decomposition of the set of factors}

\maketitle

\begin{abstract}
In this paper we explore a new hierarchy of classes of languages
and infinite words and its connection with complexity
classes. Namely, we say that a language belongs to the class $\mathcal L_k$ if it is a
subset of the catenation of $k$ languages $S_1\cdots S_k$, where the number of words of
length $n$ in each of $S_i$ is bounded by a constant.
The class of infinite words whose set of factors is in $\mathcal L_k$ is denoted by
$\mathcal W_k$. In this paper we focus on the relations between the classes $\mathcal W_k$ and the subword complexity
of infinite words, which is as usual defined as the number of factors of the word of length $n$. In particular, we prove that the class $\mathcal
 W_{2}$ coincides with the class of
 infinite words of linear complexity. On the
other hand, although the class $\mathcal W_{k}$
is included in the class of words of complexity $O(n^{k-1})$, this
inclusion is strict for $k> 2$.
\end{abstract}

\section{Preliminaries}

The complexities of infinite words and languages is a widely
studied area in formal languages theory. We follow the general approach where the complexity is measured as the number of fragments of a given size. Applied to words, it means that the complexity of a language $L$ (or an infinite word $u$) is the function $p_L(n)$ (resp., $p_u(n)$) counting the number of elements of $L$ (resp., factors of $u$) of length $n$. This function was introduced
by  Morse and Hedlund in 1938 \cite{MoHe1} under the name
\emph{block growth} as a tool to study symbolic dynamical
systems. The name \emph{subword complexity} was given by Ehrenfeucht,
Lee, and Rozenberg \cite{elr}; as the term ``factor'' replaces ``subword'', the term ``factor complexity'' is more and more popular \cite{cas_livre}.

An infinite word is ultimately periodic if and only if its complexity is ultimately constant, and it is a classical result that the smallest complexity of aperiodic words is $p(n) = n +
1$ \cite{MoHe1}. The words of this complexity are called Sturmian and form a very interesting and well-explored family (see, e.g., Chapter 2 in \cite{Lo}). Results on the complexity usually belong to one of the two families: they give either conditions or formulas on the complexity of words from given families (see, e.g., \cite{pansiot}), or conditions on words with given restrictions on the complexity. As an example of a complicated problem of that kind, we mention the $S$-adic conjecture on words of linear complexity (see \cite{leroy} and references therein). For a recent survey and deep results on subword complexity, see \cite{cas_livre}.
%for the journal paper: better survey+global complexity.

In the paper we relate the subword complexity to local conditions of factorization type.
Namely, we are interested in the following question: What is the relation between the complexity of the word and the condition that each its factor can be decomposed into a product of a finite number $k$ of words belonging to a language of a bounded complexity? In a related paper \cite{fpz} instead of languages of bounded complexity we considered the language of palindromes. 
Note that in both cases we need the language of factors to be a subset of the concatenation of these languages and not the concatenation itself. For another family of problems where the equality to the concatenation is needed, see e.g. \cite{af,hsw}.

\section{Classes and basic hierarchy}

We consider finite and infinite words over a finite alphabet
$\Sigma$, i.e., finite or infinite sequences of elements from the set $\Sigma$. A \emph{factor} or a \emph{subword} of an infinite word
is any sequence of its consecutive letters. The factor
$u_{i}\cdots u_j$ of an infinite word $u=u_1\cdots u_n \cdots$,
with $u_k \in \Sigma$, is denoted by $u[i..j]$. As usual, the set
of factors of a finite or infinite word $u$ is denoted by
Fac$(u)$. A factor $s$ of a right infinite word $u$ is called
\emph{right} (resp., \emph{left}) \emph{special} if $sa,sb \in$
Fac$(u)$ (resp., $as,bs \in$ Fac$(u)$) for distinct letters $a,b
\in \Sigma$.  The length of a finite word $s$ is denoted by $|s|$,
and the number of occurrences of a letter $a$ in $s$ is denoted by
$|s|_a$. The empty word is denoted $\varepsilon$ and we define
$|\varepsilon|=0$. An infinite word $u=vwwww\cdots = vw^{\omega}$ for some non-empty word $w$ is called ultimately ($|w|$-)periodic. In the paper we mostly follow the terminology and notation from \cite{Lo}.

%{\it (Occurrences to be distinguished from words. Short introduction on the subword complexity of words and languages (mention ult. periodic words), decompositions of languages, ???)}

Denote by $\mathcal P(\alpha)$ the set of infinite words of complexity $O(n^\alpha)$.

Let us introduce the classes $\mathcal L_k$ of languages and $\mathcal W_k$ of infinite words as follows: a language $L$ (infinite word $u$) belongs to the class $\mathcal L_k$ (resp., $\mathcal W_{k}$)
if
\[L \subseteq S_1\cdots S_k\]
(resp., Fac$(u) \subseteq S_1\cdots S_k$) for some
languages $S_i$ with $p_{S_i}(n)=O(1)$.
In other words, $u \in \mathcal W_k$ if and only if Fac$(u) \in
\mathcal L_k$, and the condition $p_{S_i}(n)=O(1)$ means exactly
that for some constant $C$ we have $p_{S_i}(n)\leq C$ for all $n$.
We also have $\mathcal P(0)=\mathcal W_1$. %S reprased

By a simple cardinality argument, we have the following inclusion:

\begin{lemma}\label{l1}
 For each integer $k>0$, we have $\mathcal W_{k+1} \subseteq \mathcal P(k)$.
\end{lemma}
\noindent {\sc Proof.} Suppose a word $u$ is in $\mathcal W_{k+1}$
and consider the factors of length $n$ of $u$. There is ${n+k
\choose k}=O(n^{k})$ ways to decompose a positive integer $n$ to
$k+1$ non-negative summands %S addends
in a given order:
$n=n_1+n_2+\ldots +n_{k+1}$. If the summand %S addend
$n_i$ is the length of the $i$th factor in a decomposition of a
word of length $n$ to $k+1$ factors, and there are at most $C$
words of length $n_i$ in the set $S_i$, it means that in total,
there are not more than $C^{k+1}$ decompositions of words
corresponding to a given decomposition of $n$. Taking all the
factors of $u$ of length $n$ together, we see that they are not
more than $C^{k+1}{n+k \choose k}=O(n^k)$, which means exactly %S the exponent k->k+1 is corrected
that $u \in \mathcal P(k)$. \hfill $\Box$

\begin{example} Now we are going to show that the Thue-Morse word $t=01101001\cdots$, defined as the fixed point starting with 0 of the morphism $\varphi: 0\to 01, 1 \to 10$, belongs to  $\mathcal W_2$. For each $n$ the Thue-Morse word consists of words $t_n=\varphi^n(0)$ and $\overline{t_n}=\varphi^n(1)$, both of them of length $2^n$: $t=t_n\overline{t_n}\overline{t_n}t_n \overline{t_n} t_n t_n \overline{t_n}\cdots$. Defining $S_1$ to be the set of suffixes of all $t_n$ and $\overline{t_n}$, and $S_2$ to be the set of their prefixes, we see that $S_1$ and $S_2$ contain exactly two words of length $k$ each. To cut each factor $w$ of $t$, we just choose any of its occurrences and a position $m$ in it divided by the maximal power $n$ of $2$: $w=t[i..j]=t[i..m]t[m+1..j]$. By the definition of $m$, $t[i..m]$ is a suffix of $t_n$ or $\overline{t_n}$, and $t[m+1..j]$ is a prefix of one of them, and thus, $w \in S_1 S_2$. So, $t\in \mathcal W_2$. This construction can be generalized to any fixed point of a 
primitive morphism but obviously not to fixed points whose complexity is higher than linear (see \cite{pansiot} for examples).
\end{example}

\begin{example} Sturmian words, which can be defined as infinite words with complexity $n+1$ for each $n$, also belong to  $\mathcal W_2$. These words have exactly one right and one left special factor of each length. One of the ways to construct the sets $S_1$ and $S_2$ for a Sturmian word $s$ is the following:
\begin{eqnarray*} S_1=\{va | a\in \{0,1\}, v \mbox{ is a right special factor of } s \} \cup \{\varepsilon\}, \\
S_2=\{av | a\in \{0,1\}, v \mbox{ is a left special factor of } s \} \cup \{\varepsilon\}.
\end{eqnarray*}
Remark that in fact the set $S_2$ is the set of reversals of factors from $S_1$, and $\# S_1 (n)=\# S_2 (n) = 2 $ for each $n>0$.
The fact that every factor of $s$ belongs to $S_1 S_2$ follows from the properties of Sturmian words: it can be proved %(we omit the details) 
that every factor $w$ of $s$ has an occurrence $[i..j]$ with $i\leq 0, j\geq 0$ in the biinfinite characteristic Sturmian word $u$ of $s$, where either $u=c^R01c$ or  $u=c^R10c$, with $c$ the right infinite characteristic word (i.e., the infinite left special word). \end{example}

Now let us introduce the {\it accumulative complexity} function %S rewritten with two formulas
$g_L(n)$ (resp., $g_u(n)$) of a language $L$ (resp., a word $u$)
as
\[g_{L}(n)=\sum_{i=1}^{n} p_{L}(n) \qquad (\mbox{resp., } g_{u}(n)=\sum_{i=1}^{n} p_{u}(n)).\]
As above, we introduce the classes $\mathcal L'_k$ of languages
and $\mathcal W'_k$ of infinite words as follows: a language $L$
(resp., infinite word $u$) belongs to the class $\mathcal L'_k$
(resp., $\mathcal W'_{k}$) if
\[L \subseteq S_1\cdots S_k\]
(resp., Fac$(u) \subseteq S_1\cdots S_k$)
for some languages $S_i$ with $g_{S_i}(n)=O(n)$.

As above, $u \in \mathcal W'_k$ if and only if Fac$(u) \in \mathcal L'_k$. The condition $g_{S_i}(n)=O(n)$ means exactly that for all $n$ we have $g_{S_i}(n)\leq Kn$ for some constant $K$.

Clearly, $\mathcal L_k \subseteq \mathcal L'_k$, since $p_{S_i}(n)\leq C$ for all $n$ implies $g_{S_i}(n) \leq Cn$. As for an opposite inclusion, we can only can prove the following theorem and its corollary.

\begin{theorem}
 $\mathcal L'_1 \subseteq \mathcal L_{2}$.
\end{theorem}
\noindent {\sc Proof.} Consider a language $L \in \mathcal L'_1$,
by definition this means  that $g_{L}(n)\leq Kn$ for some $K$. We %S rephrased
shall construct inductively the sets $S$ and $T$ of complexity
$p_S(n), p_T(n) \leq 2K+1$ such that $L\subseteq ST$.

Let us order the elements of $L$ according to their length:
$L=\{v_1,\ldots, v_n,\ldots\}$ with $|v_n|\leq |v_{n+1}|$. The
sets $S$ and $T$ are constructed inductively: we choose any
$S_1=\{s_1\}$ and $T_1=\{t_1\}$ so that $v_1=s_1t_1$ and then do
as follows. Suppose that we constructed the sets $S_{n-1}$ and
$T_{n-1}$ of cardinality less than or equal to $n-1$ each so that
$\{v_1,\ldots,v_{n-1}\}\subseteq S_{n-1}T_{n-1}$ and the number of
words of each length $l$ in each of  $S_{n-1},T_{n-1}$ is bounded
by $2K+1$.

 Consider the word $v_n$ and denote its length by $m$. It admits $m+1$
 factorizations $v_n=st$. If for a given factorization we have $s \in S_{n-1}$ and $t \in T_{n-1}$,
 we do not need to add anything to these sets and can take $S_n=S_{n-1}$, $T_n=T_{n-1}$. If for example $s \notin S_{n-1}$, we can construct $S_{n}$
 by adding $s$ to $S_{n-1}$: $S_n=S_{n-1}\cup \{s\}$ if the words of length $|s|$ in $S_{n-1}$ are at most $2K$ (and symmetrically for $T_{n-1}$).
 But the number $N$ of lengths $l$ such that $p_{S_{n-1}}(l)>2K$ (resp., $p_{T_{n-1}}(l)>2K$)
 and thus no more of words of length $l$ can be added to $S_{n-1}$ (resp., $T_{n-1}$) is bounded by $N\leq (n-1)/(2K)$,
 since the total number of words in $S_{n-1}$ (resp., $T_{n-1}$) is at most $(n-1)$.

So, to assure that at least one of $m+1$ factorizations is
admitted and we (if necessary) can add new words $s_n$ and $t_n$:
$S_n=S_{n-1}\cup \{s_n\}$, $T_n=T_{n-1}\cup \{t_n\}$ such that
$v_n=s_nt_n$, we should check that $m+1
> 2(n-1)/(2K)$. But since $m$ is the length of the word number $n$
in $L$, we have $n \leq g_L(m)\leq Km$ and thus $2(n-1)/(2K)\leq
(2Km-2)/(2K) <m+1$, which was to be proved. \hfill $\Box$

\begin{corollary}
For each $k>0$, we have $\mathcal L'_k \subseteq \mathcal L_{2k}$.
\end{corollary}
\noindent {\sc Proof.} Take a language $L \in \mathcal L'_k$: by the definition, $L\subseteq S_1\ldots S_k$ with $S_i \in \mathcal L'_1$ for all $i$. Due to the theorem above, all $S_i \in \mathcal L_2$, that is, $S_i \subseteq S_{i}^{(l)}S_i^{(r)}$ where the complexities of $S_{i}^{(l)}$, $S_i^{(r)}$ are bounded. Clearly, we have
$L\subseteq S_{1}^{(l)}S_1^{(r)}\ldots S_{k}^{(l)}S_k^{(r)}$, which proves the corollary. \hfill $\Box$

\medskip
So, for all $k>0$ we have $\mathcal L_k \subseteq \mathcal L'_k \subseteq \mathcal L_{2k}$ and thus $\mathcal W_k \subseteq \mathcal W'_k \subseteq \mathcal W_{2k}$.

\section{Linear complexity and $\mathcal W_2$}
In this section, we prove the main result of this paper, namely,
\begin{theorem}\label{t:w2} An infinite word is of linear complexity if and only if its language of factors is a subset of the catenation of two languages of bounded complexity:
 $\mathcal W_2=\mathcal P(1)$.
\end{theorem}
The $\subseteq$ inclusion has been proven in Lemma \ref{l1}.
Since for periodic words the statement is obvious, it remains
to find the languages $S,T$ of bounded complexity for a given infinite word $u$ of
linear complexity $p_u(n)\leq Cn$ such that the set of factors of $u$ is a subset of $ST$.

%In the rest of the section, we consider a non-periodic infinite word $u$ of complexity $P_u(n)\leq Cn$. Our goal is to find appropriate sets $S$ and $T$ and to prove that their complexity is indeed bounded. 
The construction of the sets  $S$ and $T$ is based on so-called {\it markers} which we define below.

\subsection{Markers and classification of occurrences}
Let $u$ be an infinite word. Given a length $n$, we say that a
subset $M$ of the set of factors of $u$ of length $n$ is a set of
{\it markers}, or, more precisely, of {\it $D$-markers} for a
constant $D$, if each factor of $u$ of length $Dn$ contains at
least one word $m \in M$ as a factor. %S reformulated

Recall that a factor $v$ of $u$ is called {\it right special} if $va, vb \in$ Fac$(u)$ for at least two different symbols $a,b$.
\begin{lemma}
The set of right special factors of $u$ of length $n$ is a set of $(C+1)$-markers, where $p_u(n)\leq Cn$.
\end{lemma}
\noindent {\sc Proof.} Consider a factor $v$ of $u$ of length $(C+1)n$ and suppose that none of its factors of length $n$ is right special. It means that each factor of $v$ of length $n$, whenever it occurs in $u$, uniquely determines the next factor of length $n$, shifted by one letter. But there are $Cn+1$ occurrences of factors of length $n$ in $v$. So, at least two of them correspond to the same factor, and what happens after its second occurrence repeats what happens after the first one. So, the word $u$ is ultimately periodic, a contradiction.  \hfill $\Box$

\medskip
The number of right special factors of $u$ of length $n$ is
uniformly bounded by a constant $R$ which is a polynomial of $C$,
where $p_u(n)\leq Cn$, due to a result of Cassaigne
\cite{cas_lin,cas_livre}. Thus, we have the following %S reformulates

\begin{corollary}\label{c:mark}
 For each length $n$, there exists a set of cardinality $R$ of $(C+1)$-markers of length $n$ in $u$.
\end{corollary}

Remark that the set of right special factors is just one the
possible ways to build the set of markers. For the proof below it
does not matter how the set of markers was constructed, the only
thing we use is that the set of markers of each length is bounded. %S reformulated

Consider a factor $w=w_1\cdots w_{n}$ of $u$ and denote by $p(w)$ its minimal period, that is,
the minimal positive integer such that $w_i=w_{i+p(w)}$ for all $i>0$ and $i+p(w)\leq n$.
The word $w[1..p(w)]$, also called the minimal period of $w$, is denoted by $P(w)$; each time it will be clear from the context whether the period means the word or the number.

An occurrence $w=u[j+1..j+n]$ of $w$ in $u$ is called {\it internal} if two conditions hold. First, $u_{j+p}=u_{j+p-p(w)}$ for all $p$ such that $1 \leq p\leq p(w)$ and $j+p-p(w)\geq 1$; second, symmetrically, $u_{j+p}=u_{j+p+p(w)}$ for all $p$ such that $n-p(w)+1 \leq p \leq n$. In other words, due to the definition of $p(w)$, for an internal occurrence of $w$ in the infinite word $u$ we have $u[j+p(w)+1..j+p(w)+n]=w$ and, provided that $j \geq p(w)$, $u[j-p(w)+1..j-p(w)+n]=w$.

An occurrence which is not internal is called {\it extreme}. More precisely, if $u_{j+i}\neq u_{j+i-p(w)}$ for some $i$ such that $\max(1,p(w)-j+1)\leq i \leq p(w)$, it is called {\it initial},
and if $u_{j+i}\neq u_{j+i+p(w)}$ for some $i$ such that $n-p(w)+1 \leq i \leq n$, it is called {\it final}. Clearly, an occurrence of a word in $u$ can be initial and final at the same time.

Since $u$ is not ultimately periodic, each its factor $w$ admits a final occurrence, otherwise $u$ would be ultimately $p(w)$-periodic.

\subsection{Construction and proof}
For each $k \geq 1$, consider the set of $D$-markers of length $2^k$ whose cardinality is bounded by $R$. Due to Corollary \ref{c:mark}, such a set exists and we shall call its elements markers of order $k$.

Consider a factor $v$ of length $n\geq 2 D$ of $u$. Our goal is to construct two words $s \in S$ and $t \in T$ such that $u=st$. By the definition of markers, $v$ contains a marker of order one; now consider the largest $k$ such that it contains a marker $m$ of order $k$. Choose an occurrence of $v$ in $u$: $v=u[i+1..i+n]$. If all occurrences of $m$ in $u[i+1..i+n]$ are internal, take one of them (say, the first one). If not, choose an extreme occurrence of $m$ in $u[i+1..i+n]$ (again, the first of them if they are several). In both cases, we denote the chosen occurrence $m=u[j+1..j+2^k]$; here $j\geq i$ and $j+2^k \leq i+n$.

Now we define $s=s(v)=u[i+1..j+2^{k-1}]$ and $t=t(v)=u[j+2^{k-1}+1..i+n]$. Clearly, $v=st$. Note that the marker $m$ is cut exactly in the middle of an occurrence: $m=m_l m_r$ with $|m_l|=|m_r|=2^{k-1}$. Here $s$ ends by $m_l$ and $t$
starts with $m_r$.

At last, let us define
\begin{align*}
S&=(\mbox{Fac}(u) \cap \Sigma^{< 2D}) \cup \{s(v)|v \in (\mbox{Fac}(u) \cap \Sigma^{\geq 2D})\}, \\
T&=\{\varepsilon\} \cup \{t(v)|v \in (\mbox{Fac}(u) \cap \Sigma^{\geq 2D})\},
\end{align*}
where $\varepsilon$ is the empty word, $\Sigma^{< n}=\bigcup_{k=0}^{n-1} \Sigma^k$ and $\Sigma^{\geq n}=\Sigma^* \backslash \Sigma^{<n}$.

It follows immediately from the definitions that Fac$(u)\subseteq ST$. It remains to prove that the cardinalities of $S \cap \Sigma^n$ and $T\cap \Sigma^n$ are uniformly bounded.

Consider a length $l\geq 2 D$. Let us count the words from $T\cap \Sigma^l$.

What can be the length of a marker $m$ used to construct a word $t \in T \cap \Sigma^l$? It is equal to $2^k$, where the word $m_r$ of length $2^{k-1}$ is a prefix of $t$ and thus $2^{k-1} \leq l$. On the other hand, since $k$ was chosen to be maximal and by the definition of $D$, we have $l<D 2^{k+1}$. These two inequalities can be rewritten as
\begin{equation}\label{e:kl}
 \frac{l}{2D}<2^k\leq 2l,
\end{equation}
which means that $k$ can take at most $\log_2D+2$ values for a given $l$.

Since we use a construction with at most $R$ markers of each order $k$, in total there are at most $R(\log_2D+2)$ markers which are used to construct the words from $T \cap \Sigma^l$. Exactly the same counting works for the words from $S \cap \Sigma^l$. They can be a bit shorter with respect to $k$ in average, since we choose the first occurrence of a longest marker whenever we have a choice, and since the factor which we decompose can be close to the beginning of $u$. However, the same bounds hold, and the same $R(\log_2D+2)$ (or less) markers can be used to construct the words from $S\cap \Sigma^l$.

Now let us consider separately the cases when the occurrence of a marker used for a decomposition is internal, initial or final.

\begin{lemma}
Consider an occurrence of a factor $v$ of length $n \geq 2D$ in $u$ and a longest marker $m$ in it. If all the occurrences of $m$ to the chosen occurrence of $v$ are internal, then $v$ is $p(m)$-periodic.
\end{lemma}
\noindent {\sc Proof.} Follows from the definition of an internal occurrence. \hfill $\Box$

Let us fix a length $l \geq 2 D$. Clearly, for a given marker $m$ of a suitable length $2^k$, there is exactly one possible word in $\Sigma^l$ which can belong to $T$ because of internal occurrences of $m$: It is $p(m)$-periodic and obtained from the prefix of length $l+2^{k-1}$ of $P(m)^{\omega}$
%define it at the beginning!
by deleting the first $2^{k-1}$ symbols. Symmetrically, there is exactly one possible word in $\Sigma^l$ which can belong to $S$ because of internal occurrences of $m$.

It follows that for each $l \geq 2 D$, each of at most $R(\log_2D+2)$ possible markers for this length, its internal occurrences can give at most one word of length $l$ in $T$ and at most one word in $S$. Now let us consider words arising from extreme occurrences.

For the sake of convenience, define a new symbol $z \notin \Sigma$ and fix $u_n=z$ for $n\leq 0$. So, instead of $u$, we can now consider a bi-infinite word $u'=\cdots zzzu_1u_2u_3\cdots$.

\begin{center}
\begin{figure}
\centering \includegraphics[width=0.8\textwidth]{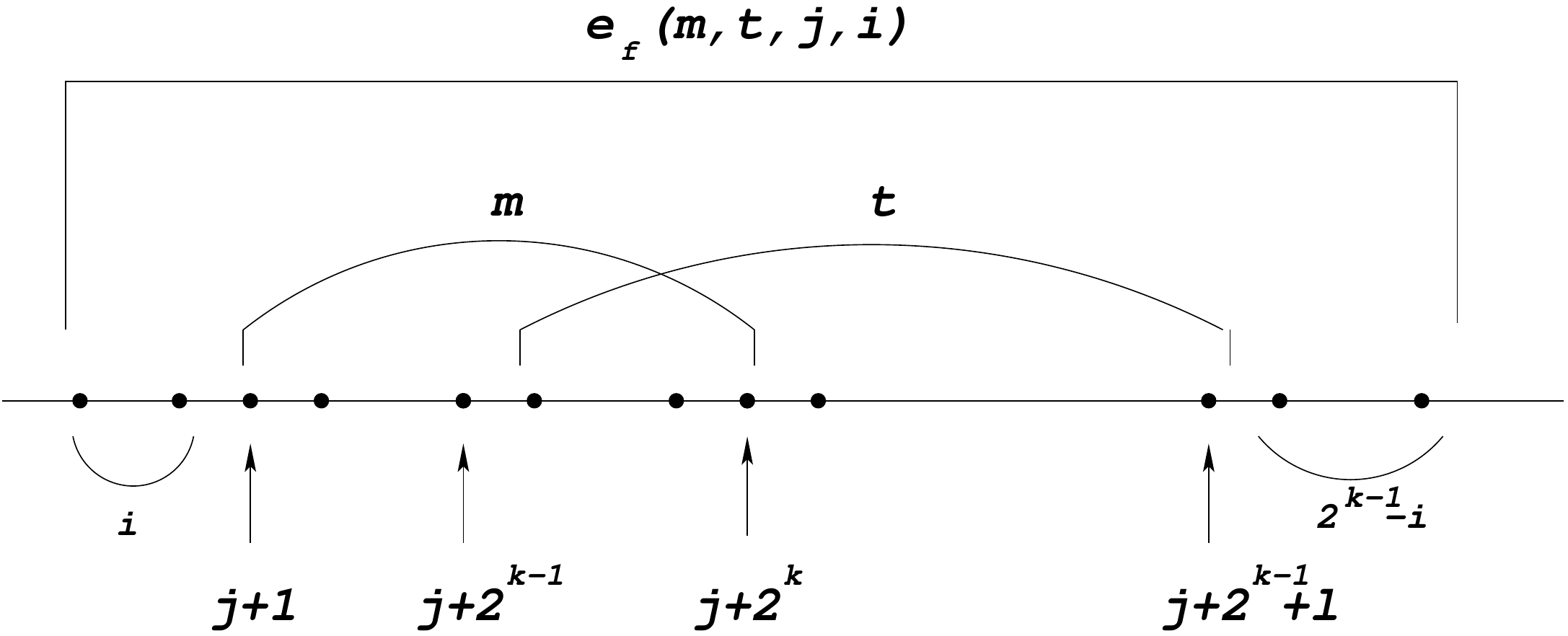}
\caption{Construction of $e_f(m,t,j,i)$}\label{f:1}
\end{figure}
\end{center}

Let us fix a marker $m$ of length $2^k$ and a length $l$ satisfying \eqref{e:kl} and consider the set $T_f(m,l)$ of words from $T$ of length $l$ arising from final occurrences of $m$ to $u$. For any word $t \in T_f(m,l)$ consider a place in $u$ which gives rise to it, that is, fix a position $j\geq 0$ such that $m=u[j+1..j+2^k]$ and $t=u[j+2^{k-1}+1..j+2^{k-1}+l]$. Now for each $i$ such that $0\leq i<2^{k-1}$ define the word $e_f(m,t,j,i)$ of length $l+2^k$ as
$e_f(m,t,j,i)=u[j+1-i..j+l+2^k-i]$ (see Fig.~\ref{f:1}). Note that if $j+1<2^k$, the word $e_f(m,t,j,i)$ for sufficiently large $i$-s starts with one or several (but not more than $2^{k-1}-1$) symbols $z$.

\begin{lemma}
 If $e_f(m,t,j,i)=e_f(m,t',j',i')$ with $|t|=|t'|=l$, then $t=t'$ and $i=i'$.
\end{lemma}
\noindent {\sc Proof.} Denote $e_f(m,t,j,i)=e_f(m,t',j',i')=e$. Note also that $k$ can be uniquely reconstructed from $m$.

Suppose that $i=i'$; then $t=t'=e[i+2^{k-1}+1..i+2^{k-1}+l]$.

Suppose that $i<i'$. Then the word $e[i+1..i'+2^k]$ has $m$ as a prefix and a suffix and thus is $(i'-i)$-periodic. In particular, $m$ is $(i'-i)$-periodic. Since $p(m)$ is the minimal period of $m$, we have $p(m)\leq i'-i <2^{k-1}=|m|/2$. So, for each $h=1,\ldots,2^k-p(m)+i'-i$ both symbols $e_{i+h}$ and $e_{i+h+p(m)}$ belong to either the prefix copy of $m$ or to the suffix copy of $m$ (or to both). So, $e_{i+h}=e_{i+h+p(m)}$ for all $h$ from 1 to $2^k-p(m)+i'-i\geq 2^k$, and in particular for all $h$ such that $2^k-p(m)+1 \leq h \leq 2^k$. This contradicts to the fact that $u[j+1..j+2^k]=e[i+1..i+2^k]$ is a final occurrence of $m$ to $u$.\hfill $\Box$

\medskip
So, the number of possible words $e_f(m,t,j,i)$ for a given marker $m$ and a given length $l$ of $t$ is minorized by the number of pairs $(t,i)$; here $t$ is a word from $T \cap \Sigma^l$ arising from a final occurrence of a marker $m$, and for each $m$, $t$ and $j$, the parameter $i$ takes exactly $2^{k-1}$ values. On the other hand, all $e_f(m,t,j,i)$ are words of length $l+2^k$, which are either factors of $u$ or its prefixes preceded by at most $2^{k-1}$ new symbols $z$: the number of factors of $u$ of length $l+2^k$ is $p_u(l+2^k)$, the number of words with $z$ is at most $2^{k-1}$, and the number of words $e_f(m,t,j,i)$ is majorized by $p_u(l+2^k)+2^{k-1}\leq C(l+2^k)+2^{k-1}$. So, we have
\[2^{k-1} t_f(m,l) \leq C(l+2^k)+2^{k-1},\]
where $t_f(m,l)$ is the contribution to $T\cap \Sigma^l $ of all the final occurrences of a marker $m$ of length $2^k$.

Since $l< 2^{k+1}D$, the latter inequality can be rewritten as
\[t_f(m,l)< \frac{C(2D+1)2^k+2^{k-1}}{2^{k-1}}=2C (2D+1)+1.\]
In other words,
\[t_f(m,l)\leq 2C (2D+1).\]

Exactly the same upper bound can be symmetrically proved for the contribution to $T \cap \Sigma^l$ of initial occurrences of a marker $m$: $t_i(m,l)\leq 2C (2D+1)$. So, each of $R(\log_2 D+2)$ possible markers for the length $l$ can contribute at most for the following number of words to $T \cap \Sigma^l$: one word arising from its internal occurrences, plus $2C (2D+1)$ words arising from final occurrences, plus $2C (2D+1)$ words arising from initial occurrences. This gives the desired upper bound: the total number of words in the set $T \cap \Sigma^l$ is bounded by the constant
\[R(\log_2 D +2)[1+4C (2D+1)].\]
The proof for $S \cap \Sigma^l$ is similar and gives the same constant as the upper bound. \hfill $\Box$

\medskip
Note that the analogous fact for general languages is not true: there exists a language of linear complexity not belonging to any $\mathcal L_k$. However, this language (which we do not describe here because of the lack of space) is not closed under taking a factor.
\section{Word of quadratic complexity}
%Lemma \ref{l1} and Theorem  \ref{t:w2} suggest the following conjecture: if $\mathcal W_2=\mathcal P(1)$ and in general, $\mathcal W_{k+1} \subseteq \mathcal P(k)$ for all $k$, is not it true that  $\mathcal W_{k+1}=\mathcal P(k)$ for all $k$?

Lemma \ref{l1} and Theorem  \ref{t:w2} imply that $\mathcal
W_2=\mathcal P(1)$, and in general $\mathcal W_{k+1} \subseteq
\mathcal P(k)$ for all $k$. So, the following natural question
arises: is it true that $\mathcal W_{k+1}=\mathcal P(k)$ for all
$k$?

The answer is negative, and, since $\mathcal W_k \subseteq \mathcal W'_k$, to show it we just point an example of a word of quadratic complexity which does not belong to $\mathcal W'_3$.

Consider the word $u=ababbabbb\cdots=\prod_{i=1}^{\infty} ab^k$. Its complexity $p_u(n)=\Theta(n^2)$: this can be either proved directly or derived from the famous paper by Pansiot \cite{pansiot}, since $u$ is obtained by erasing the first letter $c$ from the fixed point starting with $c$ of the morphism $c \mapsto cab, a \mapsto ab, b \mapsto b$.

\begin{lemma}\label{babaab}
        The word $u$ does not belong to $\mathcal W'_3$.
\end{lemma}
\noindent {\sc Proof.}
Suppose the opposite: Fac$(u)\subseteq XYZ$ with $g_X(n),g_Y(n),g_Z(n)= O(n)$. Now for each word $v \in$Fac$(u)$ of length at most $n$ fix some its decomposition $v=x(v)y(v)z(v)=xyz$ with $x \in X$, $y \in Y$, $z \in Z$. We shall estimate the number of words $v$ which can be decomposed like that.

Now for each $k,l>0$ define the word $w_{k,l}=ab^la b^{l+1} \cdots a b^{l+k-1} a$. Clearly, $w_{k,l}$ is a factor of $u$ of length $k(l+(k+1)/2)+1$.

\begin{claim}
%The number $K(n)$ of pairs  $(k,l)$ such that $|w_{k,l}|\leq n$, $k\geq 3$  and $l \geq \sqrt{n}$ is $\Theta(n \log n)$.
Let $E(n)$  be the set of pairs  $(k,l)$ such that $|w_{k,l}|\leq n$, $k\geq 3$  and $l \geq \sqrt{n}$. Then $\# E (n) = \Theta(n \log n)$.
\end{claim}
\noindent {\sc Proof.} Note that the condition $|w_{k,l}|=k(l+(k+1)/2)+1\leq n$ implies the inequality $\displaystyle l \leq \frac{n-1}{k}-\frac{k+1}{2}$. So,
\[\# E (n)= \sum_{k=3}^{\infty} \# \left \{ l \in \mathbb N: \sqrt{n} \leq l \leq \frac{n-1}{k}-\frac{k+1}{2}\right \}. \]

Observe that this set is empty for $k\geq\sqrt{2n}$: indeed, if $k\geq\sqrt{2n}$, then $\displaystyle \frac{n-1}{k}-\frac{k+1}{2} \leq \frac{n}{\sqrt{2n}}-\frac{\sqrt{2n}+1}{2}<0$. So,

$$\# E (n)= \sum_{k=3}^{\lfloor \sqrt{2n} \rfloor}
\left (\frac{n-1}{k}-\frac{k+1}{2}-\sqrt{n}+1\right ).$$
Here
 $$\sum_{k=3}^{\lfloor \sqrt{2n} \rfloor} \frac{n-1}{k} =
(n-1) \sum_{k=3}^{\lfloor \sqrt{2n} \rfloor} \frac{1}{k} = \Theta\left ( n \ln n\right )
$$
and
 $$\sum_{k=3}^{\lfloor \sqrt{2n} \rfloor} \left (\frac{k+1}{2}+\sqrt{n}-1 \right )  =
\Theta(n).$$
The claim follows. \hfill $\Box$

\medskip
Let us say that a factor $v$ of $u$ is {\it of type $(k,l)$} if $v=b^i w_{k,l} b^j$ for some $i$ and $j$. Clearly, each factor of $u$ either is of some type $(k,l)$, or contains at most one letter $a$.

Denote by $F(n)$ the set of pairs $(k,l)$ with $k \geq 3$ and $l \geq \sqrt{n}$ such that there exists a factor $v$ of $u$ of length at most $n$ and of type $(k,l)$ whose decomposition is $xyz$ with $|x|_a \leq 1$, $|z|_a \leq 1$. There were $k+1\geq 4$ letters $a$ in $v$, and at least $k-1 \geq 2$ of them stay in the word $y$. The type of $y$ is thus one of the four following: $(k,l)$,
$(k-1,l+1)$, $(k-1,l)$, $(k-2,l+1)$. But the total number of words in $Y$ of length at most $n$ is $g_Y(n)=O(n)$, and each word $y$ can give rise to at most four types from $F(n)$. So, $\# F(n)\leq 4g_Y(n)=O(n)$, and due to the previous claim, there are still $\# E(n) \backslash F(n)=\Theta(n \log n)$ pairs $(k,l)$ with $k \geq 3$ and $l \geq \sqrt{n}$ such that each word $v$ of type $(k,l)$ and of length at most $n$ is decomposed so that its middle part $y(v)$ contains at most one letter $a$. Since there are $k+1\geq 4$ letters $a$ in $v$, we see that either $x(v)$ or $z(v)$ contains at least two letters $a$.

We denote this set of pairs by $H(n)=E(n)\backslash F(n)$. The number of all factors $v$ of $u$ whose types are in $H(n)$ is denoted by $s(n)$.

Consider a factor $v$ of $u$ of length at most $n$ whose type is in $H(n)$. Suppose first that the word $x(v)$ contains more than one letter $a$. Then the word $v$ is uniquely determined by $x(v)$ and the length $|v|\leq n$. So, the number of words $v$ of length $\leq n$ admitting such a decomposition is bounded by $n g_X(n)=O(n^2)$.

Symmetrically, the number of words $v$ such that $z(v)$ contains more than one letter $a$ is bounded by $n g_Z(n)=O(n^2)$.

So, the number $s(n)$ of words whose types are in $H(n)$ is $O(n^2)$. But on the other hand, the number of types in $H(n)$ is $\Theta(n \log n)$, and for each type $(k,l)$, the number of words of this type is $l(l+k+1)$: indeed, such a word is of the form $b^iw_{k,l}b^j$, where $i$ can take $l$ values from 0 to $l-1$ and $j$ can take $l+k+1$ values from 0 to $l+k$. Since we restricted ourselves to the case of $l \geq \sqrt{n}$, the number of words of each type is $l(l+k+1)> n$. In total, we have that
$s(n) \geq n\Theta(n \log n)$, that is,
\[s(n)=\Omega(n^2 \log n).\]
A contradiction to the previous condition $s(n)=O(n^2)$. \hfill $\Box$

\medskip
Since $\mathcal W_3 \subseteq \mathcal W'_3$, we get also the following
\begin{corollary}
        There exists a word of quadratic complexity which does not belong to $\mathcal W_3$.
\end{corollary}

\section{Belonging to some $\mathcal W_k$}

The word $u$ of quadratic complexity considered in the previous section does not belong to $\mathcal W'_3$, but it can be proved that it belongs to $\mathcal W'_4$. We omit this proof here since it does not add much to the theory. However, this result suggests the following question: given a word of complexity majorated by a polynomial, is it true that it belongs to $\mathcal W_k$ for some $k$?

As we show in the next proposition, the answer to this question is
negative.

%: we build a word of complexity
%$O(n^2 f(n))$, where $f(n)$ is any slowly growing function, that
%does not belong to $\mathcal W_k$ for any $k$.

\begin{proposition}\label{n2fn} For any growing integer function $f(n)$ such that $f(1)\geq 1$, $f(n)\leq n$ and $f(n)\to \infty$, there exists an infinite
word $w$ of complexity $O(n^2 f(n))$ which does not belong to
$\mathcal W_k$ for any $k$.\end{proposition}

\noindent {\sc Proof.}
First we describe the construction of the word $w$, then we prove
that $w$ does not belong to $\mathcal W_k$ for any $k$, and after that we prove
that the word has complexity $O(n^2 f(n))$.

\medskip

Define the infinite word $w$ as follows:

$$w=\prod_{p=1}^{\infty} \prod_{q=1}^{f(p)}(a^{p}b^{q})^{k(p,q)},$$
where $k(p,q)$ is a growing function: $k(p,q)\leq k(p,q+1)$ and $k(p,f(p))\leq k(p+1,1)$ for all $p$ and $q$. %sequence of integers satisfying for all $p$ and $q$ the inequalities

%          $$ \left\{
%           \begin{array}{rcl}
%            & & k_{p,q+1} > 2 k_{p,q}, \\
%            & & k_{p+1,1} > 2 k_{p, f(p)}\frac{p+f(p)}{p+2}. \\
%           \end{array}
%           \right.$$

%These conditions ensure that each block $(a^pb^q)^{k_{p,q}}$ is at least twice longer than
%the previous block and thus longer than all the prefix before it.

Let us prove that $w\notin \mathcal W_k$ for any $k$. Suppose by contrary
that $w\in \mathcal W_k$: Fac$(w)\subseteq S_1 \cdots S_k$ with $p_{S_i}(n)\leq M_i$ for all $i$. Define $S=\cup_i S_i$; then $p_{S}(n)\leq \sum_i p_{S_i}(n)\leq \sum_i M_i = M$ for an appropriate constant $M$. Consequently, %S = substituted by \leq
$g_S(n)\leq Mn$ for all $n$.

\begin{claim}\label{c:pq}
For every pair of integers $(p,q)$, such that $p+q<\frac{n-2}{2k-1}$, $q\leq f(p)$ and $k(p,q)\geq 2k-1$, there
exists a word $s_{p,q}\in S$, $|s_{p,q}|\leq n$, such that $s_{p,q}$ contains $ba^p b^q a$
as a factor, and all those words $s_{p,q}$ are distinct.
\end{claim}

\noindent {\sc Proof.} Consider the word $b (a^p b^q)^{2k-1} a$.
Since $k(p,q)\geq 2k-1$ and $q\leq f(p)$, it is a factor of $w$,
and since $p+q<\frac{n-2}{2k-1}$, its length is at most $n$.
However we cut the word $b (a^p b^q)^{2k-1} a$ into at most $k$
pieces, at least one piece will contain $ba^p b^q a$ as a factor.
The claim follows. \hfill $\Box$

\medskip

Let us estimate the number of words $ba^p b^q a$ for $p+q<\frac{n-2}{2k-1}$,
$q\leq f(p)$ and $k(p,q)\geq 2k-1$. Since the function $k(p,q)$ is growing, there exists a constant $p_k$ such that $k(p,q)\geq 2k-1$ for all $p \geq p_k$ and all $q\leq f(p)$. Since $f(p)\leq p$ for all $p$, we have $p+q \leq p+f(p) \leq 2p$, and thus the number of pairs $(p,q)$ is bounded from below by the sum $\displaystyle \sum_{p=p_k}^\frac{n-2}{2(2k-1)} f(p)$. Since $f(p) \to \infty$ with $p$, and since $g_S(n)$ is bounded from below by the number of pairs $(p,q)$ due to Claim \ref{c:pq}, we have
\[g_S(n)\geq \sum_{p=p_k}^\frac{n-2}{2(2k-1)} f(p) > Mn\]
for some sufficiently large $n$. A contradiction to the fact that $g_S(n)\leq Mn$.

\medskip
Now let us check that the complexity of the word $w$ is $O(n^2 f(n))$. The
word $w$ contains factors of the following types:

\begin{enumerate}

\item Factors of a block $(a^{p}b^{q})^k$ for some $p$, $q$ and $k$.

\item Factors of a concatenation of blocks $(a^{p}b^{q})^{k_1}(a^{p}b^{q+1})^{k_{2}}$.

\item Factors of a concatenation of blocks $(a^{p}b^{f(p)})^{k_1}(a^{p+1}b)^{k_{2}}$.

\item Factors containing some complete block  $(a^{p}b^{q})^{k_{p,q}}$ as a factor.
\end{enumerate}
Remark that some of these families intersect, but this is not a
problem since we only need a bound. So, let us estimate the number
of words of length $n$ in each family.

In the family 1, we have $O(n)$ words of the form $a^ib^{n-i}$ or $b^ia^{n-i}$, plus $O(n^2)$ words of the form $a^i b^q a^{n-q-i}$ (uniquely determined by $0<i,q<n$) or $b^i a^p b^{n-p-i}$ (uniquely determined by $0<i,p<n$), plus words containing a factor $b a^p b^q a$ or $a b^q a^p b$. The latter words are uniquely determined by $p<n$, $q \leq f(p)$ and the position of the first occurrence of $a^p$, which takes values from 0 to $p+q<n$. So, the number of such words (and thus of all the words in family 1) is $O(n^2f(n))$.

Treating the other three families analogously, we see that the complexity of each of them is at most $O(n^2f(n))$ too.
\iffalse
In the family 2, a word not belonging to family 1 contains factors $a b^q a$ and $b^{q+1}$. So, such a word is uniquely determined by $p<n$, $q \leq f(p)$ and the position of the first occurrence of $b^{q+1}$, which takes values from 0 to $n-q-1<n$. So, the number of such words is also $O(n^2f(n))$.

An analogous counting works for the family 3: the new words
appearing there are determined by $p<n$ and the first position of
$a^{p+1}$, so, their number is $O(nf(n))$.

At last, for the family 4, let us choose the first complete block $(a^{p}b^{q})^{k(p,q)}$ in a given word. Then this word is determined by $p$, $q$ and the position of its beginning, so, the number of such words is again $O(n^2 f(n))$.
\fi
So, the complexity $p_w(n)=O(n^2f(n))$, which completes the proof. \hfill $\Box$

\section{Conclusion}

We finalize this paper by suggesting the following open problem: What is the minimal possible complexity of a word which does not
belong to any $\mathcal W_k$?

Remark that Theorem \ref{t:w2} and Proposition \ref{n2fn} imply
that this complexity is strictly bigger than linear and is at most
quadratic.

\medskip
%\section*{Acknowledgement}
Supported in part by RFBR grants 12-01-00089 and  12-01-00448, as well as by the Academy of Finland grant 251371.

\end{document}